\documentstyle[aps,twocolumn,epsf]{revtex}

\begin{document}

\draft

\title{A stochastic rainbow caustic observed with cold atoms}

\author{B.T. Wolschrijn, D. Voigt, R. Jansen, R.A. Cornelussen, N.
Bhattacharya, R.J.C. Spreeuw and H.B. van Linden van den Heuvell}

\address{Van der Waals-Zeeman Institute, University of Amsterdam, \\
         Valckenierstraat 65, 1018 XE Amsterdam, the Netherlands\\
         e-mail: spreeuw@wins.uva.nl}

\date{\today}
\maketitle

\begin{abstract}

We report the direct observation of a novel type of rainbow
caustic. In contrast to known examples, this caustic originates
from a dissipative, stochastic process. We have observed this
using cold {$^{87}$Rb} atoms bouncing inelastically on an
evanescent-wave atom mirror. The caustic appears as a sharp peak
at the lower edge of the asymmetric velocity distribution of the
bouncing atoms. The stochastic process is a spontaneous Raman
transition due to photon scattering during the bounce. The results
are in good agreement with a classical calculation.

\end{abstract}



Caustics are ubiquitous phenomena in nature. Examples are the
cusp-shaped patterns of light reflection on the inside of a coffee-cup and
the patterns of bright lines observed on the bottom of a swimming pool
\cite{Ber80}. The prime example of a caustic is the common rainbow, which
can be understood in a ray-optics picture by considering how the scattering
angle of a light ray depends on its impact parameter on a water droplet
\cite{Nus77} . Whereas the incident rays have smoothly distributed impact
parameters, the outgoing rays pile up where the scattering angle has a
local extremum. Such a divergence of the ray density, the caustic, appears
at the rainbow angle. In atomic \cite{ForWhe59} and nuclear \cite{BraSat97} scattering
experiments analogous rainbow phenomena have also been observed.

The examples of caustics that have been known so far have in common that
the outgoing parameter (scattering angle) is a {\em deterministic} function of
the incoming parameter (impact parameter). In this Letter we report on
our observation of a new type of rainbow caustic
existing by virtue of a {\em stochastic} process, which distributes a single-valued
``impact parameter'' over a range of ``scattering
angles''.
To our knowledge, such stochastic caustics have not been observed before.

We have observed this caustic in the vertical velocity distribution of cold atoms,
after bouncing inelastically off an evanescent-wave mirror
\cite{SodGriOvc95,DesArnSzr96,LarOvcBal97,OvcManGri97}.
Previous experiments on the transverse velocity distribution of
atoms bouncing elastically on corrugated mirrors
\cite{LanCogHor97,RosHalHug00} also allow an interpretation in terms of caustics.
However, those were of the usual deterministic kind, where the outgoing
transverse velocity is a deterministic function of the position where the
atom hits the mirror. The caustic then originates from atoms reflecting from
inflection points on the mirror surface.

Whereas other experiments on inelastically bouncing atoms
concentrated on the cooling properties, the appearence of the
velocity caustic seems to have escaped attention. The incident
atoms in our experiment are nearly monochromatic, i.e. they have a
narrow velocity distribution $\Delta v/v=0.03$. During the bounce
the atoms are optically pumped to a different hyperfine
ground-state, by a spontaneous Raman transition. They leave the
surface with less kinetic energy, due to the difference in optical
potential. The outgoing atoms have a broad velocity distribution.
This is due to the stochastic nature of the spontaneous Raman
process, so that atoms make the transition at different depths in
the evanescent wave. Our measurements and analysis show that the
resulting velocity distribution is highly asymmetric. The caustic
appears as a sharp peak at the minimum possible velocity. This
divergence demonstrates the strong preference for making the Raman
transition at the turning point.

The experiment is performed in a rubidium vapor cell. We trap
about $10^{7}$ atoms of {$^{87}$Rb} in a magneto-optical trap
(MOT) and subsequently cool them in optical molasses to
{15~$\mu$K} corresponding to a r.m.s. velocity spread of
$\sigma_v=3.8$~cm/s. The MOT has a $1/e^{2}$ radius of $0.6$~mm
and is centered $5.8$~mm above the horizontal surface of a
right-angle BK7 prism. When the falling cloud arrives at the
surface, the mean velocity is $34$~cm/s, and the velocity spread
is $\sigma_v=0.9$~cm/s. Note that this distribution is for fixed
$z=0$, not for fixed time. At the surface, we create an evanescent
wave (EW) by a Gaussian shaped laser beam of $0.8$~mm $1/e^{2}$
radius, which undergoes total internal reflection.

The blue detuned EW induces a repulsive optical dipole potential
$U_{F}(z)=U_{F}(0) \exp(-2\kappa z)$, where the subscript $F=1,2$
denotes the hyperfine ground state,
$\kappa=k_0\sqrt{n^2\sin^{2}\theta-1}$, with $k_{0}=2\pi/\lambda$
the free space wave vector of the light, $n=1.51$ the refractive
index and $\theta$ the angle of incidence
\cite{CooHil82,BalLetOvc87,VoiWolSpr00}. The maximum potential
$U_{F}(0) \propto I_{0}/\delta _{F}$, where $I_{0}$ is the
incident intensity at the glass surface, and $\delta _{F}$ is the
detuning. Here we define the detuning $\delta _{1,2}$ relative to
the transition $5S_{1/2}(F=1,2)\rightarrow 5P_{3/2}(F'=F+1)$ of
the $D_2$ line ({780 nm}), see Fig.\ref{expscheme}. Note that we
neglect the van der Waals surface attraction. This is justified as
long as $U_{F}(0)$ is much larger than the kinetic energy of the
atoms. The main influence of this force is a reduction of the
effective mirror area \cite{LanCouLab96}.

An atom entering the EW in the $F=1$ state is slowed down by the potential
$U_{1}(z)$. Spontaneous Raman-scattering can transfer the atom  to the higher
hyperfine ground state ($F=2$).
When transferred into this state, the potential acting on the atom is
$U_{2}(z)$, which is
weaker due to the increase of the detuning by approximately
the hyperfine splitting $\delta_{2} \approx   \delta_{1} + \delta_{\rm hfs}$
(Fig.\ref{expscheme}).
The ratio of the two potentials, $\beta \equiv
U_{2}(0) / U_{1}(0) $, quantifies the reduction in potential energy.
As a result the atoms will bounce inelastically,
i.e. to a lower height than their initial release position.

Experimental data of bouncing atoms are taken by absorption imaging.
After a variable time delay the atomic cloud is exposed to
{$20~\mu$s} of probe light, resonant with the
$5S_{1/2}(F=2)\rightarrow 5P_{3/2}(F'=3)$ transition.
Thus only atoms which have been transferred to $F=2$ contribute to the signal.
The atomic cloud is imaged on a digital frame-transfer CCD camera.
The presented images show an area of {$2.2\times4.5~$mm$^2$} with a pixel
resolution of {$15~\mu$m} (Fig.\ref{figResults}) .
The initial position of the MOT is outside the field of view. The
horizontal line at the bottom shows the prism surface.

A typical series with {2~ms} time increments  is shown in
Fig.\ref{figResults}, where $t=0$ is defined as the time that the
center of the cloud reaches the mirror. Each density profile has
been converted into a horizontal line sum. The solid curve is the
result of a calculation described below. The amplitude of the
experimental curves is rescaled such that the maximum optical
density of the experimental curve coincides with the theoretical
maximum value. This is the only fit parameter.

As expected, the atoms bounce up less high than their initial
height. Furthermore, the spatial distribution shows another striking
feature: it displays a high density peak at low $z$,
and a long low-density tail extending upward.
Note also that there is a time-focus: the density peak is sharpest
when the slowest atoms reach their upper turning point.
This density peak is an immediate consequence of a caustic
appearing at the minimum possible velocity.

The caustic can be understood by considering the atoms as point
particles moving in the EW potential. This corresponds to the
ray-optics limit for the optical rainbow. We consider an atom
arriving at the surface in its $F=1$ hyperfine ground state, with
an initial downward velocity $v_{i}<0$. Its trajectory through
phase space is determined by energy conservation:
$U_{1}(0)\exp(-2\kappa z) + \frac{1}{2}mv^{2} =
\frac{1}{2}mv_{i}^{2}$, and is depicted in Fig.\ref{figcaustic} by
the thick line. While it is slowed down by the EW it may scatter a
photon at a velocity $v_{p}$, and be transferred to the $F=2$
state. The atom continues on a new trajectory determined by
$U_{2}(z)$. To illustrate the formation of the caustic, possible
trajectories starting at various positions in phase space are
depicted as thin gray ($v_{p}<0$) and black ($v_{p}>0$) curves.
For asymptotically large $z$ the density of curves represents the
outgoing velocity distribution, showing the caustic where the
trajectories pile up. This distribution is similar to a rainbow.
The density of outgoing trajectories diverges at the
\emph{"rainbow velocity"}. Below this velocity the intensity is
zero, similar to Alexander's dark band in the optical rainbow
\cite{Nus77}. Above the rainbow velocity the intensity
distribution decreases smoothly.

Despite the similarities, there is a crucial difference between a
rainbow created by sunlight refracted by waterdroplets and our
'velocity caustic'. The appearance of a rainbow is due to a {\it
deterministic} process, where the scattering angle is uniquely
determined by the impact parameter. In our experiment the incoming
velocity $v_{i}$ is nearly single-valued, and is distributed over
a broad range of outgoing velocities by the {\em stochastic}
process of spontaneous Raman scattering.

Given the novel character of this caustic, the analogy with known
examples can only be valid to some extent. For example, the
position of the caustic is independent of the size parameter
$\kappa$. Similarly the optical rainbow angle does not depend on
the droplet size. We measured for three different values of the
decay length $\kappa^{-1}$ the atomic density profile at the upper
turning point (Fig.\ref{fig:resultbeta}a). This is defined as the
highest position reached by the peak density. The shape of the
cloud did not change, but the number of atoms decreased with
decreasing $\kappa^{-1}$ since the total number of scattered
photons is lower.

The limitation to the analogy becomes apparent through the parameter $\beta$ which determines
the position of the caustic.
This parameter characterizes the ``degree of dissipation'' and therefore has no analogy in
the optical rainbow, or any other deterministic caustic.
For example, the optical rainbow angle is determined by the index of refraction $n$ of water.
However, the energy of the photons is not changed, and therefore $n$ cannot be compared to $\beta$.
The change in the position of the caustic with $\beta$ can clearly be seen in Fig.\ref{figcaustic}b,c.
Here we measured the height of the upper turning point, as a fraction of the initial MOT height,
for three values of the detuning.

In order to quantitatively analyze our experimental data, we write
the optical hyperfine-pumping rate during the bounce as
$\Gamma'(z)=(1-q)\Gamma U_{1}(z)/\hbar \delta_{1}$, where $q$ is
the branching ratio to $F=1$. We define $\eta(v)$ as the survival
probability for the atom to reach the velocity $v$ without
undergoing optical pumping. This function decreases monotonically
as $\dot{\eta} = - \Gamma ' \eta$, with $\eta(v_{i})=1$. For
$|v|\leq |v_{i}|$, the solution is $\eta(v) =
\exp(-(v-v_{i})/v_{c})$ , where $v_{c} \equiv  2 \kappa \hbar
\delta_{1} /(1-q)m\Gamma$, with $m$ the atomic mass. When the
pumping process takes place at a certain velocity $v=v_{p}$, the
atom leaves the surface with velocity
$v_{f}=\sqrt{v_{p}^{2}(1-\beta )+\beta v_{i}^{2}}$
(Fig.\ref{expscheme}). This results in a distribution of bouncing
velocities, resulting from atoms which were pumped while moving
toward ($v_{p}^{-}$) or away from ($v_{p}^{+}$) the surface :
$n(v_{f}) = \eta(v_{p}^{-}(v_{f})) \times
|\frac{\partial{v_{p}^{-}}}{\partial{v_{f}}}| +
\eta(v_{p}^{+}(v_{f})) \times
|\frac{\partial{v_{p}^{+}}}{\partial{v_{f}}}|$ for
$\sqrt{\beta}|v_{i}| \leq v_{f} < |v_{i}|$ , which diverges at
$v_{f}=\sqrt{\beta}|v_{i}|$. This velocity caustic originates from
atoms which are pumped near the turning point. There are two
reasons for scattering preferentially at this position. An atom
spends a relatively long time at the turning point since its
velocity is lowest there. In addition, the intensity of the EW,
and thus the photon scattering rate is highest at the turning
point. The divergence is an artefact of the ray-optics
description. It disappears due to diffraction when the atoms are
treated as matter waves.

To compare the experimentally obtained spatial distributions with
the model, we first calculated the one-dimensional phase-space
density $\rho(z,v)$. The spatial distribution is obtained by
projecting $\rho$ on the $z$-axis. Initially the MOT is described
by a normalized gaussian, $\rho(z,v)\propto \exp(-(z-z_{0})^{2}/2
\sigma_{z}^{2})\exp(-v^{2}/2\sigma_{v}^{2})$ with $z_{0}=5.8$~mm,
$\sigma_{z}=0.3$~mm and $\sigma_{v}= 3.8$~cm/s. The cloud falls
due to gravity and expands due to thermal motion. When it arrives
at the EW, its velocity distribution is nearly gaussian centered
around $v=34$ cm/s, with a spread of $\sigma_{v}=0.9$ cm/s. When
the cloud is reflected, its velocity distribution is changed in
the way described above. We measure the spatial distribution of
inelastically bounced atoms after a time of flight $t$. To
illustrate the agreement with the experimental data, the result
for a time $t=7$~ms is drawn as the thick line in
{Fig.\,\ref{figResults}.

A very intriguing aspect of our experiment is the possibility to
observe supernumeraries: interference of two trajectories with the
same outgoing velocity, but with opposite pumping velocity
$v_{p}$. In the velocity distribution after the inelastic bounce,
the high velocity tail is the sum of two contributions: from atoms
moving towards and moving away from the surface when they are
pumped to $F=2$. Interference between these two paths should
result in oscillations in the velocity distribution. This is a
nontrivial effect because it involves the spontaneous emission of
a photon.

Given our experimental parameters, we expect a typical oscillation period of
$1$~cm/s.
This spacing depends on $\kappa$, just as in a rainbow the supernumeraries
depend on the droplet size.
These oscillations are not yet resolved in our measurements, because in the EW-field the
magnetic substates are not degenerate, each producing a different velocity
distribution.

In conclusion, we have observed a new type of caustic due to the {\it
stochastic} distribution of a monochromatic input.
The caustic appears as a sharp peak at the lower edge of the velocity
distribution of rubidium atoms, bouncing inelastically on an evanescent-wave atom mirror.
The caustic bears some resemblance to the common optical rainbow.
However its dissipative character makes it uncomparable with any deterministic caustics
and qualifies it as a new physical phenomenon.

This work is part of the research program of the ``Stichting voor Fundamenteel
Onderzoek van de Materie'' (FOM)
which is financially supported by the ``Nederlandse Organisatie
voor Wetenschappelijk Onderzoek'' (NWO).
R.S. has been financially supported by the Royal Netherlands
Academy of Arts and Sciences.

\bibliographystyle{prsty}

\begin{figure}
  \centerline{\epsfxsize=8.0cm\epsffile{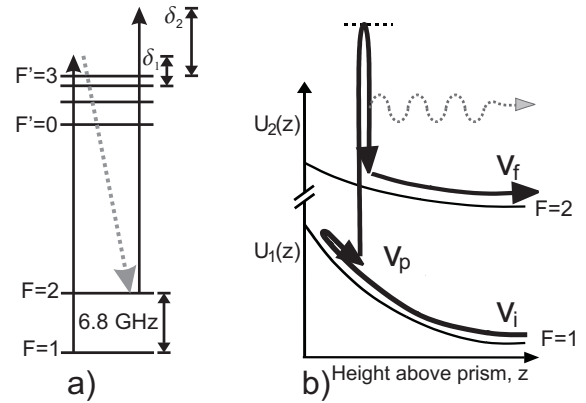}}
  \vspace*{0.5cm}
  \caption{
    (a) Relevant atomic levels of {$^{87}$Rb}
    (b) An atom enters the evanescent wave in its $F=1$ state
with initial velocity $v_{i}$. It is decelerated and spontaneously scatters a
photon at a velocity $v_{p}$. After being pumped to $F=2$
it accelerates and leaves the potential with asymptotic velocity $v_{f}$. }
\label{expscheme}
\end{figure}

\begin{figure}
  \centerline{\epsfxsize=8.0cm\epsffile{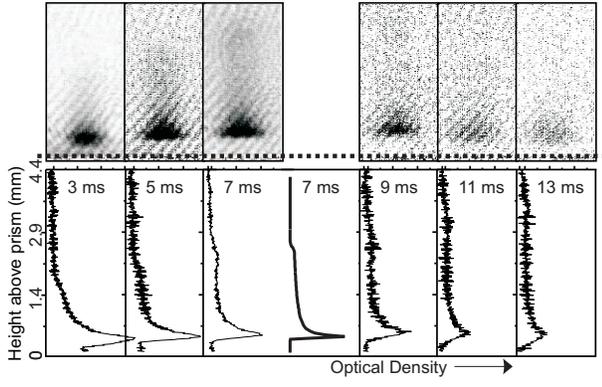}}
  \vspace*{0.5cm}
  \caption{
    Absorption images ($2.2\times 4.5$mm$^{2}$) at different moments
after bouncing on an EW, with detuning $92 \Gamma$, and $\kappa^{-1}=1
\mu$m, where $t=0$ is defined as the time that the center of the cloud
reaches the mirror.
 A line sum shows the atomic
density distribution above the prism.
The dotted line indicates the prism
surface. The initial MOT was located at 5.8 mm above the surface.
The thick curve in the middle is the result of our calculation at $7$ ms.}
\label{figResults} \end{figure}

\begin{figure}
  \centerline{\epsfxsize=8.0cm\epsffile{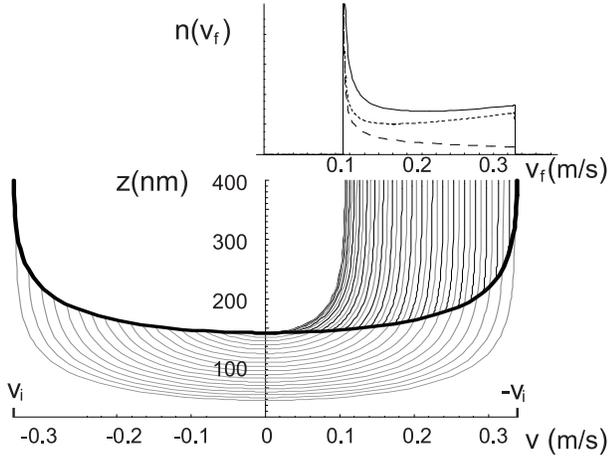}}
  \vspace*{0.5cm}
  \caption{
Construction of the velocity caustic in terms of phase space
trajectories. The velocity of the incident atoms is $v_{i}=-0.34$
m/s. The thick solid line shows the trajectory of the lower
hyperfine state bouncing elastically. If the atom is pumped to the
other hyperfine state it continues on a different trajectory. The
thin lines represent possible outgoing paths, each starting at a
different pumping velocity, depending on the position of
Raman-transfer. The density of outgoing trajectories diverges,
yielding a caustic in the velocity distribution. Shown in the
upper curve are the total distribution (solid line), and the
contribution of atoms moving towards (dotted) and away from
(dashed) the surface while being transferred.} \label{figcaustic}
\end{figure}

\begin{figure}
  \centerline{\epsfxsize=8.0cm\epsffile{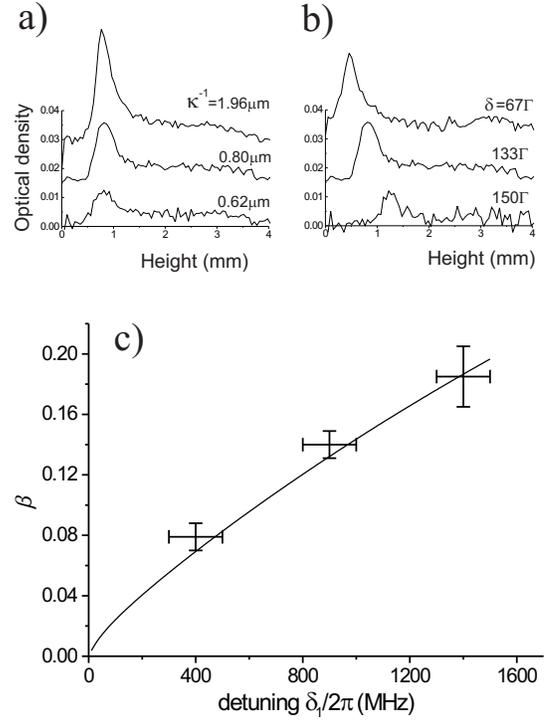}}
  \vspace*{0.5cm}
  \caption{  a) Density
profiles at $t=14$ ms after the bounce for $\kappa^{-1}= 1.96 \mu$m, $0.80 \mu$m
and $0.62\mu$m, $\delta_{EW}=150 \Gamma$. The curves have been offset by 0.015. Variation of $\kappa$ in
this range, has no influence on the position and shape of the spatial
density profile.   b) Density profiles for various detunings: $\delta_{EW} =
67, 150$ and $233 \Gamma$, again with 0.015 offset. The curves are measured at the moment that the
density peak reaches its highest point and have the same scale.
    c) Measured height of the upper turning point
as a fraction of the initial (MOT) height. This represents
$\beta$, the relative optical potential strength of the $F=1$ and $F=2$
ground states. The curve results from a calculation including the full
level structure.}
\label{fig:resultbeta}
\end{figure}


\end{document}